# Delayed Ultrafast X-ray Auger Probing (DUXAP) of Nucleobase Ultraviolet Photoprotection

B. K. McFarland[1], J. P. Farrell[1,2], S. Miyabe[1], F. Tarantelli[3], A. Aguilar[4], N. Berrah[5], C. Bostedt[6], J. Bozek[6], P.H. Bucksbaum[1,2], J. C. Castagna[6], R. Coffee[6], J. Cryan[1,2], L. Fang[5], R. Feifel[7], K. Gaffney[1], J. Glownia[1,2], T. Martinez[1,8], M. Mucke[7], B. Murphy[5], A. Natan[1], T. Osipov[5] , V . Petrovic[1,2], S. Schorb[6], Th. Schultz[9], L. Spector[1,2], M. Swiggers[6], I. Tenney[1,2], S. Wang[1,2], W. White[6], J. White[1,2] and M. Gühr[1*]

[1] PULSE, SLAC National Accelerator Laboratory, Menlo Park, CA 94025, USA
[2] Departments of Physics and Applied Physics, Stanford University, Stanford, CA 94305, USA
[3] Dipartimento di Chimica, Universita di Perugia, and ISTM-CNR, 06123 Perugia, Italy
[4] Advanced Light Source, Lawrence Berkeley National Lab, Berkeley, CA 94720, USA
[5] Physics Department, Western Michigan University, Kalamazoo, MI 49008, USA
[6] LCLS, SLAC National Accelerator Laboratory, Menlo Park, CA 94025, USA
[7] Department of Physics, Uppsala University, Uppsala, Sweden
[8] Department of Chemistry, Stanford University, Stanford, CA 94305, USA
[9] Max-Born-Institut, 12489 Berlin, Germany
*mguehr@slac.stanford.edu

## Abstract

We present a new method for ultrafast spectroscopy of molecular photoexcited dynamics. The technique uses a pair of femtosecond pulses: a photoexcitation pulse initiating excited state dynamics followed by a soft x-ray (SXR) probe pulse that core ionizes certain atoms inside the molecule. We observe the Auger decay of the core hole as a function of delay between the photoexcitation and SXR pulses. The core hole decay is particularly sensitive to the local valence electrons near the core and shows new types of propensity rules, compared to dipole selection rules in SXR absorption or emission spectroscopy.

We apply the delayed ultrafast x-ray Auger probing (DUXAP) method to the specific problem of nucleobase photoprotection to demonstrate its potential. The ultraviolet photoexcited $\pi\pi^*$ states of nucleobases are prone to chemical reactions with neighboring bases. To avoid this, the single molecules funnel the $\pi\pi^*$ population to lower lying electronic states on an ultrafast timescale under violation of the Born-Oppenheimer approximation. The new type of propensity rule, which is confirmed by Auger decay simulations, allows us to have increased sensitivity on the direct relaxation from the $\pi\pi^*$ state to the vibrationally hot electronic ground state. For the nucleobase thymine, we measure a decay constant of 300 fs in agreement with previous quantum chemical simulations.

## Introduction and Motivation

Molecules have the ability to convert the energy of sunlight into other degrees of freedom like heat, electric current or store it in altered chemical bonds. These processes require the involvement of multiple electronic states of the molecule and therefore proceed in violation of the Born-Oppenheimer approximation (BOA) [Born_1927] that allows separation of the electronic and nuclear degrees of freedom. Often, these excited state processes and nonadiabatic effects due to BOA breakdown occur on an ultrafast timescale, ranging from femtoseconds to picoseconds [Yarkony_1996, Domcke_2004, Levine_2007]. The failure of the BOA has triggered a series of investigations, combining quantum chemistry on photoexcited molecular states and ultrafast experiments on isolated molecules, to advance our understanding of nature's light harvesting machines. On the experimental side, ultrafast spectroscopy with two or more femtosecond pulses is used to investigate the molecular path. The first (pump) pulse, in the visible or ultraviolet spectral range, excites the molecular electrons (see left hand side of Fig. 1) giving way to a coordinated motion of electrons and nuclei. A second time-delayed (probe) optical light pulse is used to report on the evolution of the electrons and the nuclei. A number of different types of probe pulses can be used, and a particularly useful scheme has been time-resolved photoionization spectroscopy, where the probe pulse ionizes the molecule and either ion yield [Kotur_2011, Kotur_2012] or the kinetic energy of the ejected electrons [Stolow_2004] is measured.

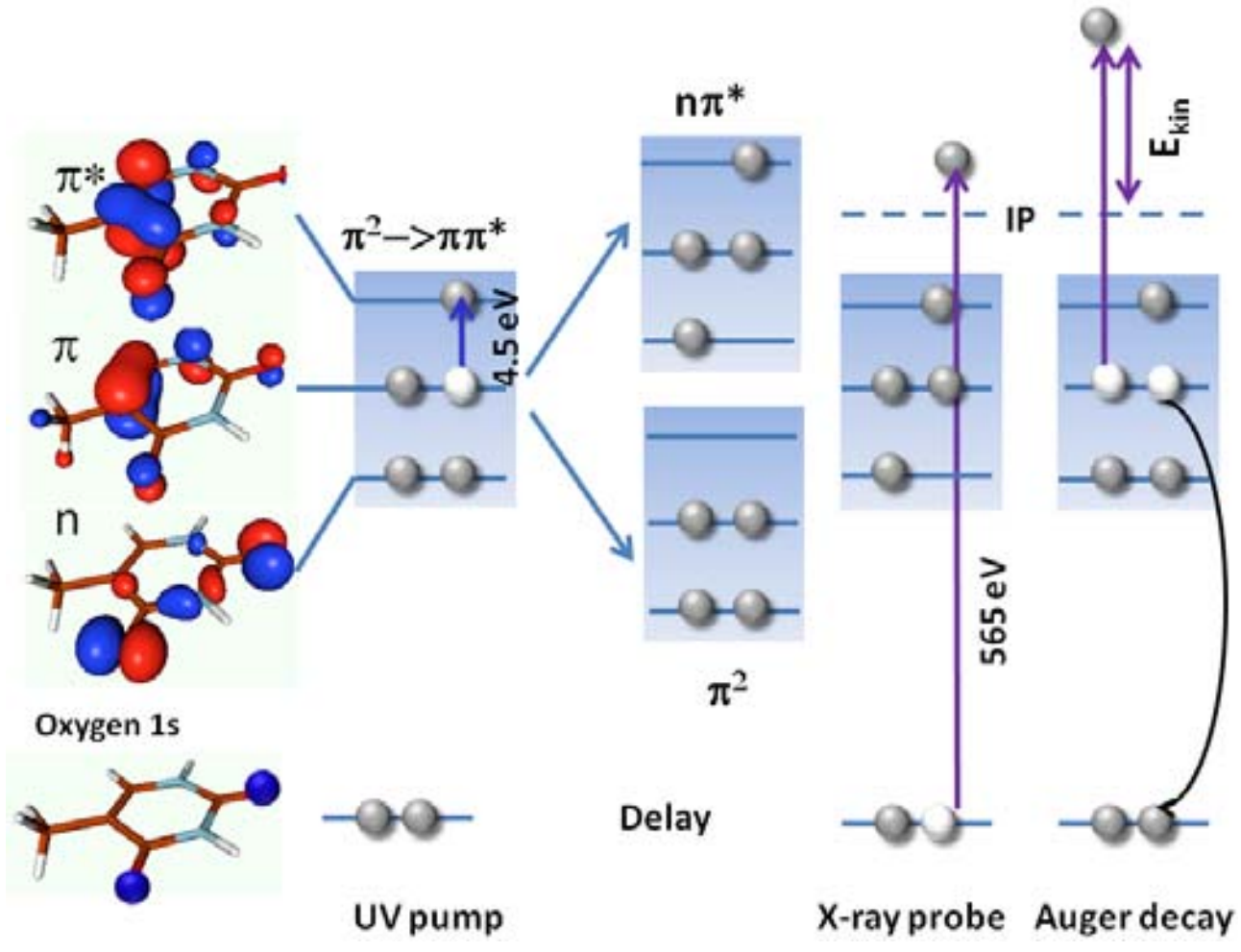

**Fig. 1:** DUXAP (delayed ultrafast x-ray induced Auger probing) scheme illustrated with the example of the thymine photoprotection scheme. The images on the right side show the *n*,*π* and *π** molecular valence orbitals as well as the oxygen 1s orbitals, where the core hole can be created. The *π* and *n* orbitals are both doubly occupied in the ground state configuration. The UV pump pulse promotes an electron from a *π* orbital to a *π** orbital, giving rise to the chemically reactive *ππ** state. The time scales and yields for competing electronic relaxation to the ground state $\pi^2$ or *nπ** or pure nuclear relaxation of the *ππ** state are currently a matter of debate. We probe the electronic valence state by a delayed soft x-ray (SXR) probe pulse creating a core hole and inducing Auger decay. The emitted Auger electrons carry information about the valence electron state at the position of the core hole (oxygen in our case).

We introduce a new experimental method where the probe is a soft x-ray (SXR) pulse. This induces core ionization, which is promptly followed by Auger decay (see Fig. 1 right hand side). We analyze the energy dispersed Auger electrons as a function of delay between the ultrashort UV pump and SXR probe pulses. This spectroscopic method, which we call delayed ultrafast x-ray induced Auger probing (DUXAP), exhibits several advantages as a method to investigate the (non-BOA) nuclear dynamics occurring on the time scale of the pump-probe delay:

a) The binding energies of the core electrons are distinct for different atoms inside the molecule and the Auger process probes mostly the chemical environment of the core-ionized atom. Thus, the Auger decay in a selected energy window displays highly local properties of the valence electronic states.
b) While the vibrational relaxation can inhibit ionization with ultraviolet probe pulses, the soft x-rays are capable of core ionizing the molecule in any nuclear geometry.

c) We discover new types of propensity rules that are *opposite to dipole selection rules* and thus allow a different view on the valence electronic states.
d) The Auger decay is insensitive to strong spectral fluctuations typical for a self amplified spontaneous emission (SASE) soft x-ray source, which is currently the most common free electron laser source [Emma_2004].

In application to dilute quantum systems the DUXAP method complements the now established method of ultrafast x-ray absorption spectroscopy applicable to high density targets [Bressler_2004]. The necessary short x-ray pulses at relatively high pulse energy have become possible due to the development of the linac coherent light source (LCLS) [Emma_2010]. For the future more of these sources are planned or in construction so that the method pioneered here can find further application and even extension to larger systems.

We demonstrate the power of DUXAP on the example of ultrafast nucleobase photoprotection. Nucleobases, encoding genetic information inside DNA, are strong absorbers in the ultraviolet range transmitted by the earth's atmosphere. After UV excitation, the nucleobases are placed in highly reactive electronically excited states, which can lead to their dimerization on an ultrafast timescale [Schreier_2007]. The observed small quantum yield for UV induced chemistry is thought to be due to competing ultrafast relaxation out of the electronically excited states into the ground state [Crespo_2004]. It has been speculated that the five common nucleobases, all showing this unusual (for aromatic molecules) ultrafast electronic decay, have been evolutionarily selected for their photostability given their crucial task inside DNA [Kang_2002].

The precise mechanism for electronic decay of the light induced excited state is a matter of ongoing discussion. The absence of excited state fluorescence was the first indication for an ultrafast decay to the ground state [Daniels_1971]. A new era of nucleobase photoprotection studies were triggered by the progress of femtosecond laser systems, as documented in the review by Kohler *et al.* [Crespo_2004]. In general, all studies measure excited state decay rates in the femtosecond to picoseconds regime for nucleobases in liquid environments and isolated nucleobases in the gas phase. The ultrafast nature of the measured transients hints at a radiationless decay mechanisms occurring via the BOA violation. This observation confirms the importance of non-BOA dynamics and conical intersections (CIs) in general [Yarkony_1996, Domcke_2004, Levine_2007]. In this paper we concentrate on the nucleobase thymine which has only one dominant structural isomer, avoiding mixed spectra from several isomers.

The important valence orbitals of thymine are plotted on the left of Fig. 1. In the electronic ground state, the $\pi$ and $n$ orbitals are doubly occupied. Ultraviolet light promotes an electron from the $\pi$ orbital into the initially unoccupied $\pi^*$ orbital, resulting in a chemically reactive $\pi\pi^*$ state [Schreier_2007]. Ab initio quantum chemical studies of isolated thymine point towards several possibilities for the ultrafast decay from $\pi\pi^*$ towards the electronic ground state, in which the molecule is completely protected from dimer formation. A direct femtosecond decay from $\pi\pi^*$ directly into the vibrationally hot ground state was suggested as the dominant deexcitation channel through an ethene like conical intersection [Perun_2006, Merchan_2006, Asturiol_2009]. In addition, an indirect decay from $\pi\pi^*$ to the spectroscopically dark $n\pi^*$ (doubly occupied $\pi$, singly occupied $n$ and $\pi^*$ orbitals) has been suggested. The relative weight of this decay to a dark excited $n\pi^*$ state compared to direct decay to the ground state, as well as the corresponding rates, has been a matter of debate, even on the theoretical level [Hudock_2007, Szymczak_2009, Asturiol_2009, Asturiol_2010]. The relaxation from spin singlet to the triplet states is expected to occur on a 10 ps timescale for thymine [Etinski_2009]. While the ultrafast nature of the process is confirmed by first principles dynamics calculations, the ratio of indirect to direct decay and the timescale for the direct decay remains a matter of discussion.

Several experimental studies in liquids and isolated molecules confirm that the excited state decay occurs in the sub-picosecond and picosecond regime [Pecourt_2000, Kang_2002, Ullrich_2004, Canuel_2005,

Hare_2007, Gonzales-Vasquez_2009]. The isolated studies are best suited to be compared to the single molecule *ab-initio* theory discussed above. Time resolved ion yield experiments in the gas phase show a very short sub 100 fs decay constant followed by additional 5-7 ps decay [Kang_2002, Canuel_2005, Gonzales-Vasquez_2009]. The short decay has been interpreted either as a coherent experimental artifact [Kang_2002] or a real decay to the electronic ground state in agreement with the theoretical prediction [Canuel_2005]. Photoemission experiments on isolated molecules add a third decay constant in the 500 fs regime [Ullrich_2004], which was interpreted to display the population trapping in the $\pi\pi^*$ minimum [Hudock_2007]. Other authors attribute the 500 fs time constant to an erroneous fitting model [Canuel_2005]. Given the essential theoretical disagreement about direct ground state decay and indirect $n\pi^*$ decay and the experimental dispute about analysis and interpretation of transients, it will be advantageous to measure the ground state refilling after UV excitation.

So far, there is no indication of an experimental signal reflecting the isolated molecule ground state. The high molecular densities in liquid environments allow for ground state transient absorption experiments [Hare_2007] showing a sub-picosecond direct (thought to occur from $\pi\pi^*$) and an indirect many ten picosecond channel via the $n\pi^*$ and triplet states. To compare theory and experiment directly, we attempt to measure the transient population of the ground state using soft x-ray probe pulses.

**Experimental setup**

Our experiments are performed at the LCLS at SLAC National Accelerator Laboratory and at the Advanced Light Source (ALS) at Lawrence Berkeley National Laboratory. We first describe the setup at LCLS in Fig. 2.

Our experiment takes place in the high field physics (HFP) vacuum chamber at LCLS [Bozek_2009]. The molecules are evaporated in an oven at about 140 deg C and guided through a 900 micron capillary of about 40mm length to the photon interaction region. Measurements on the beam ex situ delivered a sample density of about $10^{11}$ molecules/cm$^3$ at the exit of the capillary and a molecular beam opening angle in the range of 10 degrees. The background pressure was around $10^{-9}$ mbar, corresponding to particle densities that are 3-4 orders of magnitude below the thymine density in the interaction region. The molecules are first irradiated by a UV pulse of 266 nm wavelength having a duration of 70 fs. After a time delay in the femtoseconds to picosecond domain, the photoexcited molecules are exposed to a soft x-ray pulse of 565 eV photon energy with an estimated duration from 50 to 80 fs. The SXR pulses have a spectral jitter and bandwidth of about 0.25-0.5%. The relative delay between UV and SXR pulse jitters from shot to shot. We characterize the jitter using a single shot, x-ray optical cross-correlator to measure and re-sort the shots according to delay, giving 70 fs accuracy [Schorb_2012]. The UV pump pulse was focused to a ~70 micron spot estimated from a microscope pointing at a YAG screen in the interaction region. The x-rays were focused more tightly than this. The energy of the UV pulses and the x-ray pulses were chosen to avoid nonlinear effects, as checked by the spectra upon a systematic variation of both pulse energies independently.

The SXR pulse creates 1s core holes in carbon, nitrogen and oxygen inside the molecule according to their relative absorption cross sections. The photoelectrons are discarded, since their energy fluctuates strongly with the LCLS spectral jitter. Instead we analyze the Auger electrons around 500 eV kinetic energy, which are exclusively ejected from an O 1s core hole decay. The electrons are funneled towards a 2 m long flight tube by a strong permanent magnet. They encounter a retardation of 470 V and hit an MCP detector to measure their time of flight (TOF) which is then converted to kinetic energy. The spectrometer was designed as a magnetic bottle [Frasinski_2013]; however we turned the solenoid in the flight tube off to increase the spectral resolution. With this setup we were able to distinguish lines spaced by 2 eV in the Xe-MNN-Auger spectrum [Werme_1973], which also serve as our calibration standard. The polarizations of the UV and SXR pulses are aligned with the flight tube symmetry axis.

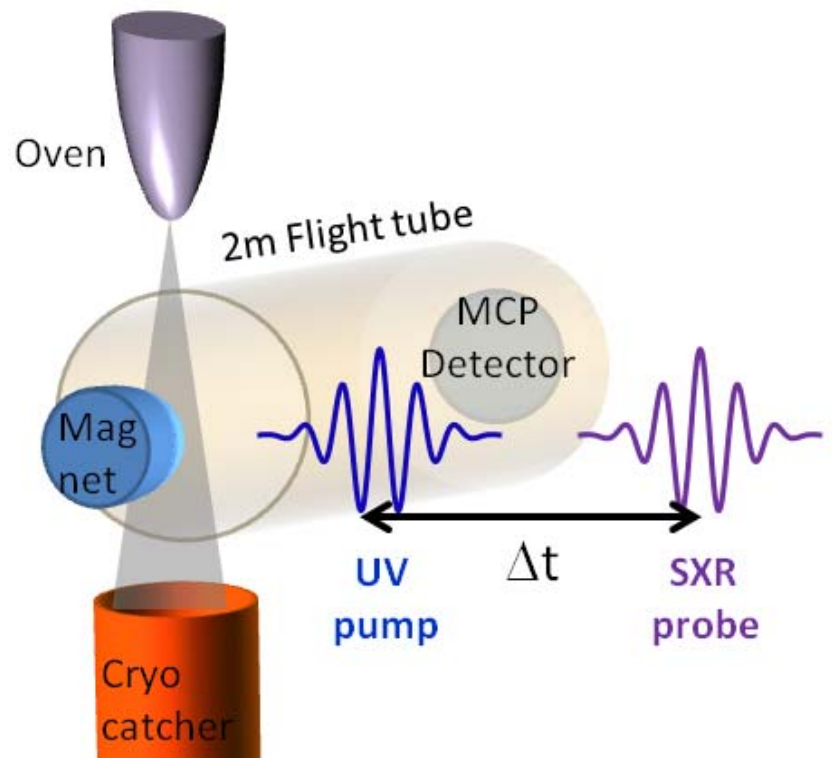

**Fig.2**: Experimental setup at the LCLS. The molecules are sprayed out of a heated oven into the interaction region which is inside a magnetic bottle spectrometer consisting of magnet, a 2m long flight tube and a multi channel plate (MCP) detector. The molecules are photoexcited by the ultraviolet (UV) pump pulse. After some delay, the soft x-ray probe pulse core ionizes the molecule. We resolve the Auger decay of the oxygen core hole as a function of kinetic energy and delay between pump and probe pulse.

At the ALS, we used beamline 8 to record Auger spectra of unexcited molecules at variable photon energy and for N, C and O core hole decay. The kinetic energy of the Auger spectra was dispersed in a hemispherical analyzer whose axis was aligned to 56 deg with respect to the soft x-ray polarization. The molecules were evaporated at similar temperatures to LCLS into a gas cell flanged on the analyzer.

## Theoretical Methods

The Auger spectra of the ground state molecule, at its equilibrium geometry (Fig. 3) and at a variety of excited state stationary point geometries (see appendix), are computed by the second-order Algebraic Diagrammatic Construction, ADC(2), method [Schirmer_1984, Tarantelli_2006]. This yields the spectrum of double ionization energies which are then subtracted from the core ionization potential to obtain the Auger electron energies. The intensities of the Auger transitions are estimated by a two-hole population analysis of the ADC(2) eigenvectors, according to the method outlined previously [Tarantelli_1991]. Spectra such as that in Fig. 3 are then obtained by Gaussian broadening of the discrete lines (full-width-at-half-maximum 1.5 eV). The calculations have been performed using the cc-pVDZ basis set [Dunning_1989].

To calculate the Auger spectra of excited electronic states, we pursue a different strategy outlined in the following. We define the total Auger decay rate for production of dication in configurations i and the outgoing Auger electron with energy $E$ by

**Equation 1:** $$\gamma(E) = \sum_{i=1}^{\substack{\text{Dication}\\ \text{Configuration}}} c_i^2 I_i(E)$$

where $c_i$ is the coefficient vector determined by the complete active space-configuration interaction (CASCI) calculation on the dication (see below for more detail), and $I_i$ is the Auger decay rate. We compute $I_i$ using single determinant wavefunctions describing the initial core-hole state ($\Phi_0$) of the ion and final state ($\Psi_{\mathbf{k}_i}^{(-)}$) with outgoing momentum $\mathbf{k}_i$ for the Auger electron. In this case the Fermi Golden rule expression for the total Auger decay rate with energy normalization of the final state is

**Equation 2:** $$I_i(E) = \frac{2\pi}{\hbar} \int \left| \left\langle \Psi_{\mathbf{k}_i}^{(-)} \middle| \sum_{i \neq j} \frac{e^2}{r_{ij}} \middle| \Phi_0 \right\rangle \right|^2 d\Omega_{\mathbf{k}}$$

where $d\Omega_{\mathbf{k}}$ is the solid angle of the Auger electron. There are two cases that must be considered. In the first case the Auger electron and the electron that repopulates the core-hole comes from the same molecular orbital. In the second case the two electrons are from different molecular orbitals. Siegbahn gives the detailed derivation of the expressions [Siegbahn_1975]. Following his treatment, the rate for the first case is

**Equation 3:** $$I_i(E) = \frac{2\pi}{\hbar}\int\left|\left\langle \varphi_c(1)\varphi^{(-)}{}_{\mathbf{k}_i}(2)\left|\frac{e^2}{r_{12}}\right|\varphi_v(1)\varphi_v(2)\right\rangle\right|^2 d\Omega_{\mathbf{k}}$$

where $\varphi_c$ and $\varphi_v$ are the core and valence molecular orbitals, respectively. The continuum orbital for the Auger electron, $\varphi^{(-)}_{\mathbf{k}_i}$ , is computes using the electron-molecule scattering wavefunction obtained with the complex-Kohn method [Rescigno_1995] at the static-exchange level of theory (for details see [Petrovic_2012]). The scattering calculations were carried out using Dunning's double-zeta basis [Dunning_1970]. Note that the decay rates are computed using the molecular orbitals from a Hartree-Fock calculation on the dication. Eq. 3 is just a standard two-electron repulsion integral and the rate is proportional to $|J|^2$, where $J$ is the Coulomb integral. If the residual ion of the final state is singlet coupled, and the decay process involves two different valence orbitals, the rate is

**Equation 4:**

$$I_i^{Singlet}(E) = \frac{2\pi}{\hbar}\int\frac{1}{2}\left|\left\langle \varphi_c(1)\varphi^{(-)}_{\mathbf{k}_i}(2)\left|\frac{e^2}{r_{12}}\right|\varphi_v(1)\varphi_{v'}(2)\right\rangle + \left\langle \varphi_c(1)\varphi^{(-)}_{\mathbf{k}_i}(2)\left|\frac{e^2}{r_{12}}\right|\varphi_v(2)\varphi_{v'}(1)\right\rangle\right|^2 d\Omega_{\mathbf{k}}$$

where $\varphi_v$ and $\varphi_{v'}$ emphasize the fact that two different valence orbitals are involved in the transition. This expression has a form $\frac{1}{2}|J+K|^2$, where $K$ is the exchange integral. For the triplet case, the rate is given as:

**Equation 5:**

$$I_i^{Triplet}(E) = \frac{2\pi}{\hbar}\int\frac{3}{2}\left|\left\langle \varphi_c(1)\varphi^{(-)}_{\mathbf{k}_i}(2)\left|\frac{e^2}{r_{12}}\right|\varphi_v(1)\varphi_{v'}(2)\right\rangle - \left\langle \varphi_c(1)\varphi^{(-)}_{\mathbf{k}_i}(2)\left|\frac{e^2}{r_{12}}\right|\varphi_v(2)\varphi_{v'}(1)\right\rangle\right|^2 d\Omega_{\mathbf{k}}.$$

which can again be expressed in terms of Coulomb and exchange integrals as $I_i^{Triplet}(E) = \frac{2\pi}{\hbar}\int\frac{3}{2}|J-K|^2 d\Omega_{\mathbf{k}}$.

The kinetic energy of the Auger electron is determined by the energy difference between the initial core-hole state and the final bound residual dication states. The absolute energy differences between the core-hole state and the ground dication state at various molecular geometries are determined by the ADC(2)] method explained above. Note that we have shifted the computed energies so that the oxygen 1s ionization potential of neutral thymine at equilibrium geometry matches the experimental value [Plekan_2008]. We then carry out the CASCI(26/15)/6-311G* calculation on the ground and excited states of the dication to determine the relative kinetic energy of the Auger electron along with the final state configurations and $c_i$ used to compute the total Auger decay rate (Eq. 1). To be consistent with the ADC(2) approach and our scattering calculations we used Hartree-Fock (HF) molecular orbitals in the CASCI calculation. For the ground state spectrum in the energy range of our interest (505eV~520eV) this approach gave good agreement with the two-hole population analysis of the ADC(2)/cc-pVDZ calculation explained above (see Fig. 1 in Appendix 1).

The experiment tracks the nuclear wavepacket dynamics on the ground and excited state surfaces of thymine. Thus, we computed the Auger decay rates of neutral thymine in various electronic states (ground, $\pi\pi^*$, and $n\pi^*$ states) at key geometries previously determined by Hudock *et al.* [Hudock_2007] that included Franck-Condon, $\pi\pi^*$ minimum, $n\pi^*$ minimum and ($\pi\pi^*/n\pi^*$) conical intersection geometries. We are also interested in the "hot ground state spectrum" since the wavepacket created on the $\pi\pi^*$ state eventually returns to the ground state surface with extra energy. To compute such a spectrum we ran a classical trajectory for 200 fs on the ground state potential energy surface computed using HF/6-311G* method starting at the (ground/$\pi\pi^*$) conical intersection geometry determined by Asturiol *et al.* [Asturiol_2009]. We then randomly picked five geometries from the dynamical calculation to compute the averaged Auger electron spectrum.

**Results and Discussion**

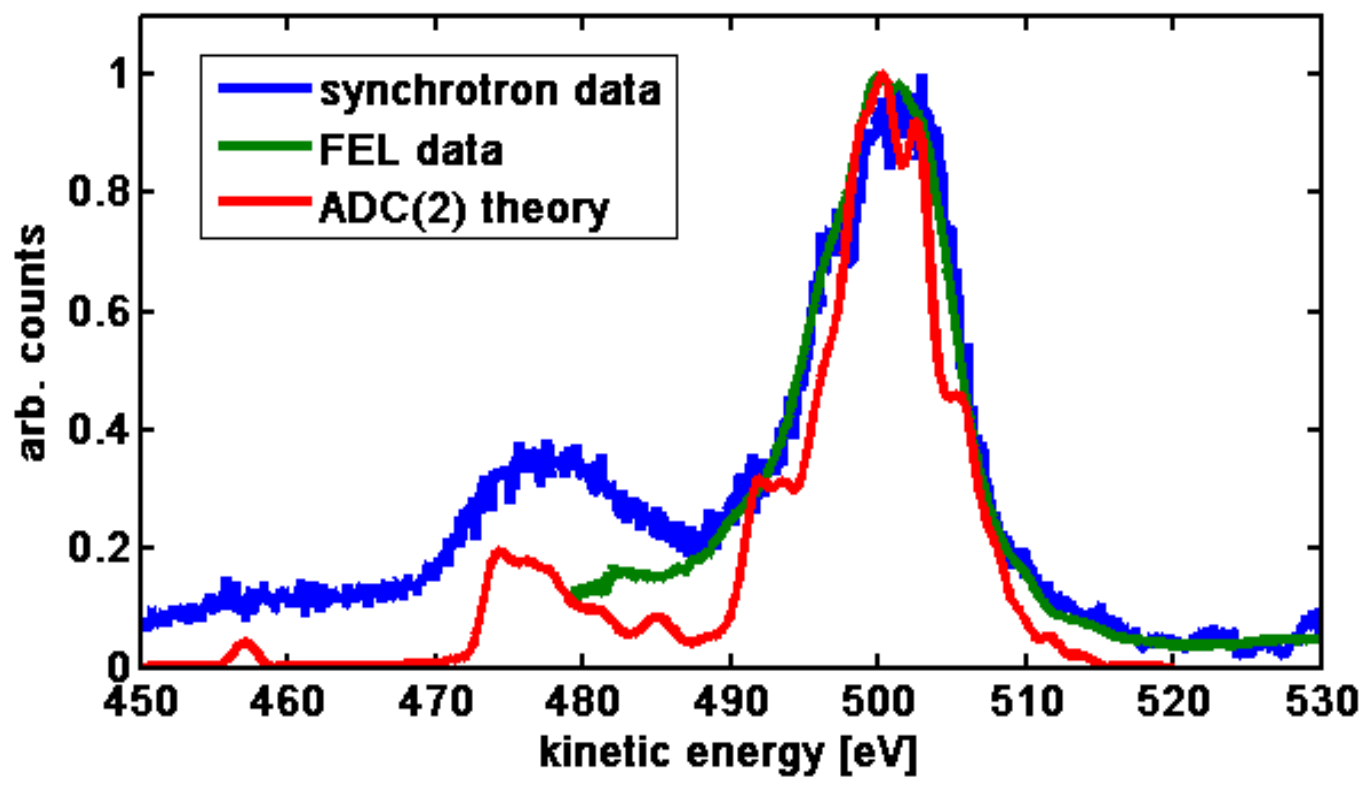

**Fig. 3:** Comparison of experimental and theoretical ground state Auger spectra for thymine. The blue and green curves are data taken at the advanced light source synchrotron and the LCLS free electron laser, respectively. The magnetic bottle spectrometer at the FEL used a retardation voltage of 470 eV leading to a reduction in transmission for the lower kinetic energy range until about 487 eV. The ADC(2) calculations (red) fit the data very well.

We first present the Auger spectrum of the unexcited thymine molecule. The green curve taken at the LCLS starts only at 480 eV, since we used a retardation potential of 470 eV to optimize the energy resolution of the instrument. The ALS data extends over a larger range and shows more features in agreement with the ADC(2) calculations shown in red. The photon energy used at the ALS was 546 eV; at LCLS it was 565-570 eV which is considerably higher than the ionization limit of 540 eV at the oxygen site [Plekan_2008]. The kinetic energy at the LCLS was carefully calibrated with the Xe Auger features, the ALS data were shifted to lower kinetic energies by 1 eV, which is due to a less involved calibration. It is remarkable that the ADC peak positions overlap that well with the data.

The different Auger bands at 500, 475 and 457 eV involve deeper valence orbitals with lower kinetic energy. In the high energy region from about 500 to 520 eV, the most dominant contribution from the decay comes from n orbitals, which are lone pair orbitals strongly localized at oxygen. The highest energy lone pair orbital is depicted in Fig. 1. We can intuitively understand why the Auger decay from n is so intense. The Auger matrix element (Eq. 3 and 4) connects the core hole state with the final state having two holes in the valence manifold. The spatial overlap of core hole and valence holes thus is important for the Auger amplitude. We calculate two types of strong Auger decays: one involving two electrons from the highest bound n orbital according to Eq. 3, the second involving one electron of the highest bound and a second electron from a deeper bound orbital also strongly localized at the oxygen atoms. The latter Auger amplitude is evaluated according to Eq. 4. Appendix 1 describes the details of the decay patterns.

This finding has important consequences for the Auger decay of the various states after photo excitation. The $\pi\pi^*$ state has two n electrons similar to the ground state and we expect similar Auger amplitudes and similar energies of the final dicationic states. The $n\pi^*$ state however has one electron missing in the n orbital which is contributing to all strong Auger lines. The first decay pattern according to Eq. 3, with both electrons from the n orbital, is thus forbidden. The decay energies of the second type Auger decay according to Eq. 4 are shifted to much lower kinetic energies due to the interaction of the two n holes with the deeper lying n like valence hole (see Appendix 1 for more detail). The region between 500 and 520 eV kinetic energy, in the focus of our study, is therefore insensitive to the $n\pi^*$ state. Based on some simple energy considerations, the $n\pi^*$ state Auger decay might be shifted into the energy range below 500eV (see end of Appendix 1).

a)

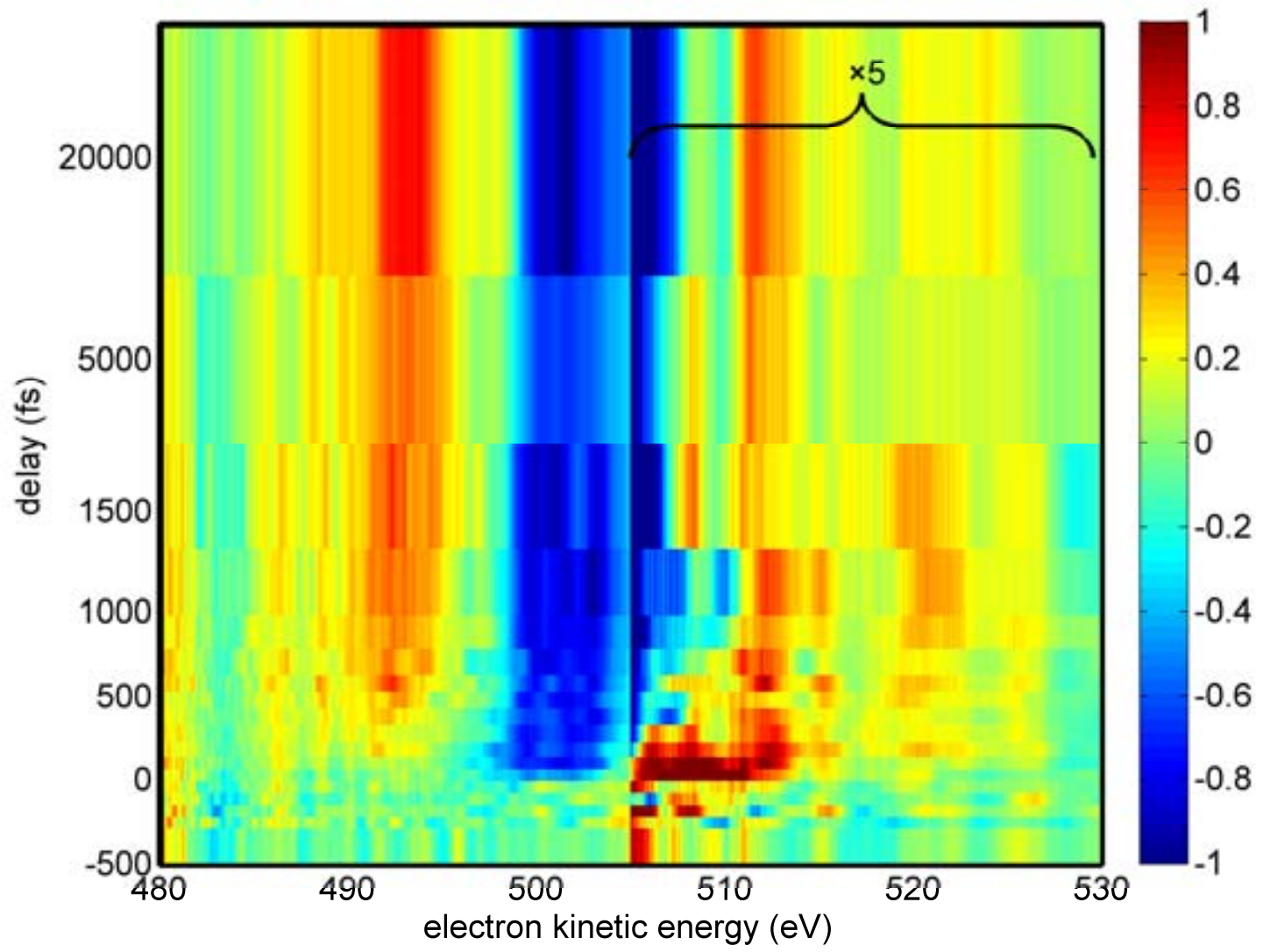

b)

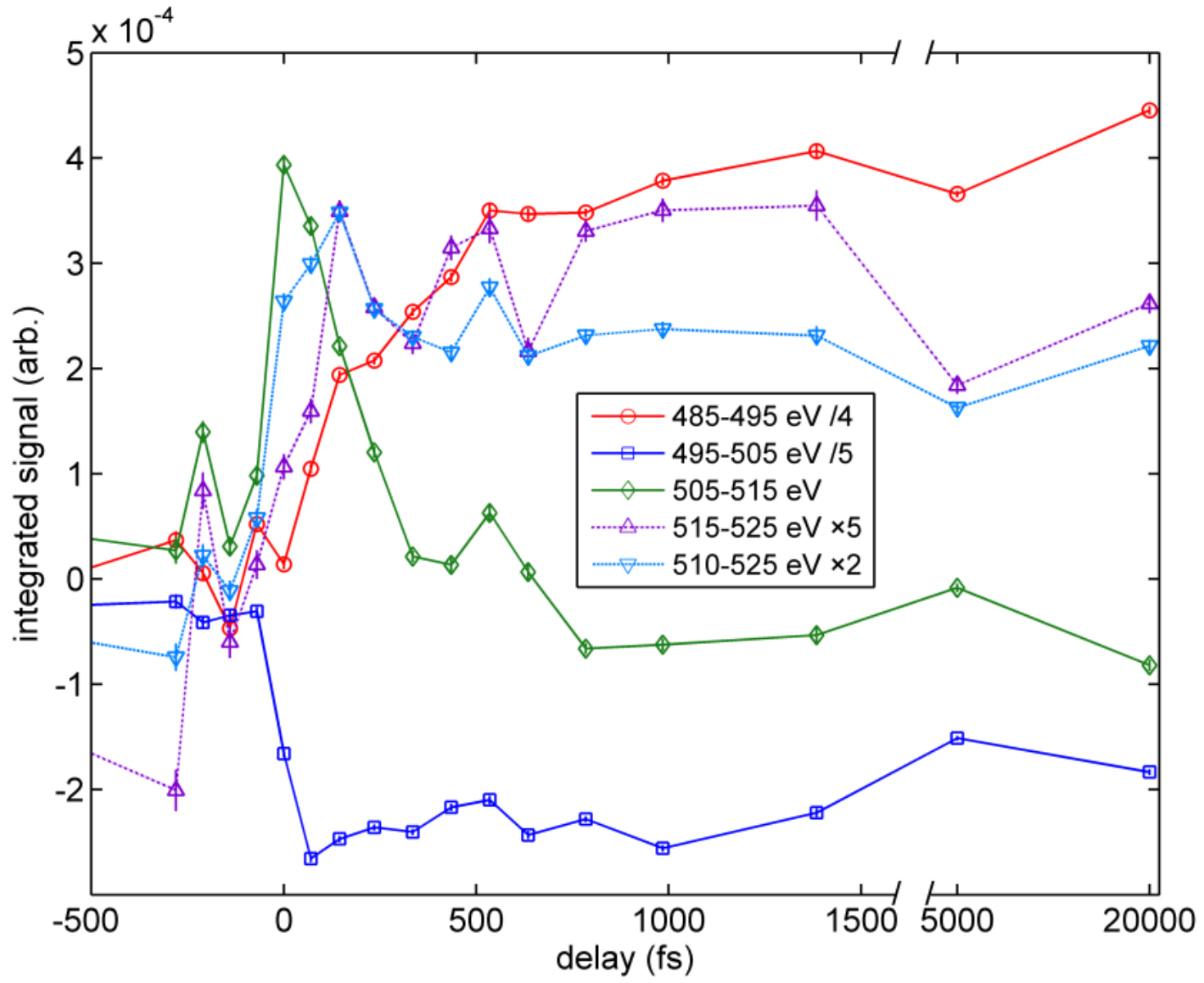

**Fig.4**: a) Difference Auger spectra D(delay,$E_{kin}$)=Signal($UV_{on}$)-Signal($UV_{off}$) as function of Auger electron kinetic energy and time delay between UV pump and soft x-ray probe pulse, color code: blue: UV induces less Auger decay, red: UV induces more Auger decay. The region between 505 and 530 eV is scaled by a factor of 5 for better visibility. b) Integrations over kinetic energy range in a) as given in the legend.

Figure 4a shows the color coded spectral amplitude difference between Auger spectra with and without ultraviolet excitation. The blue regions represent a UV induced reduction of the Auger decay, whereas the red regions indicate a UV enhanced Auger decay. The spectra are plotted as a function of pump probe delay with non-equidistant time intervals. The time zero was determined from the fastest features within the spectrum. Before zero delay no systematic features are visible in the transient Auger spectrum. The difference spectrum reflects directly the difference between the unexcited and excited molecule. Assuming a fraction *f* of the molecules are excited, the difference signal is:

**Equation 6:**

$$D(delay, E_{kin}) = Signal(UV_{on}) - Signal(UV_{off}) =$$
$$Signal_{GS}(1-f) + Signal_{ES} f - Signal_{GS} = f(Signal_{ES} - Signal_{GS}),$$

where the subscripts *GS* and *ES* refer to the ground and photoexcited molecular states, respectively.

The Signal($UV_{on}$) and Signal($UV_{off}$) spectra are constructed from individual shots, that fall into a selected delay interval. We evaluate the signal by binning the single shots randomly and analyzing a histogram of the added signal heights of the binned spectra. The Signal is set to be the mean of this histogram, the error is the standard error of this mean typically varying from 10 to 30%. For further discussion, we divide the transient spectrum into several kinetic energy zones according to the features visible and plot the integrated signal in these zones as a function of delay in Fig. 4b. Since the spectrum is essentially independent of time after 1.5 ps, we concentrate on early delays in part b. The error bars are estimated from the errors of the individual energy resolved spectra. The fastest feature visible in Fig. 4b is signal decay in the region between 495 and 505 eV and rise in the 505 to 515 eV kinetic energy regions. The distance between measured points in delay is 70 fs. The 505-515 eV channel shows a transient maximum at short delays before it decays in about 300 fs. The rise in the 515-525 eV window is clearly delayed to the 505-515 eV window rise by 140 fs. The 485-495 eV channel is most delayed compared to all other integration windows and shows an increase to its full magnitude after 500 fs.

In the following we attribute these channels to different electronic states of the molecule. The fastest signals must be related to the depletion of the ground state and the creation of an initial wave packet in the $\pi\pi^*$ excited state. The electronic ground state of thymine has its strongest Auger features in the spectral range between 495-505 eV, exactly where we observe the fastest transient. Thus we attribute this channel to ground state bleaching, which stays constant up to 20 ps delay between UV and soft x-ray pulses. This means that the initial nuclear geometry of the ground state is not reached even 20 ps after the photo excitation. Furthermore, the bleach seems to be the only feature constant over the full 20 ps delay range with negligible change. We therefore deduce that the spectral region between 495 and 505 eV is free of Auger decay channels from photoexcited states. From the bleach amplitude to unexcited amplitude ratio of 10% we can deduce that we are in the weak field limit with the UV excitation. The immediate rise of the signal in the 505-515 eV channel must be related to the newly born wave packet on the only allowed electronic transition to the $\pi\pi^*$ state. We therefore deduce that these decay energies are particular sensitive to the $\pi\pi^*$ state and reflect mostly nuclear coordinates in the Franck-Condon range with the ground state. The subsequent signal decay on the order of 300 fs reflects the wave packet propagation on the $\pi\pi^*$ towards the crossings with other electronic states. We now turn to very long decay times of 20 ps which, besides the bleach, also shows a signal increase in the 485-495 eV and 510-525 eV regions. From 1.4 to 20 ps delay the spectrum does not change within the experimental uncertainty. This indicates the occupation of long lasting states whose identity we determine. The rise time of the difference signal in this spectral window varies from prompt between 505 and 515 eV to delayed by 150 fs between 515 -525 eV. Moreover, we observe a rise and decay behavior on both windows between 0 and 300 fs indicating a transient population. The final steady signal, and thus the final electronic state, is reached around 300 fs.

The candidates for long lasting states in the literature are clearly the vibrationally hot ground state [Perun_2006, Merchan_2006, Asturiol_2009] and the $n\pi^*$ [Hudock_2007, Szymczak_2009, Asturiol_2010] or a mixture of both. As explained above, we expect very little Auger decay at high kinetic energies from the $n\pi^*$ state. In the difference spectrum, this would result in a negative (blue) feature at long time. In contrast we see a red feature indicating that the state populated must give rise to more Auger decay than the ground state in this energy window according to Eq. 6. The calculated Auger difference spectra shown in Fig. 5 confirm our qualitative arguments. Indeed the $n\pi^*$ decay is much weaker than the GS decay, which suggests the observation of the hot vibrational ground state. We rule out the triplet from a comparison with gas phase calculations [Etinski_2009], predicting a time constant for its population of about 10ps in agreement with liquid phase triplet observations [Hare_2007].

a)

b)

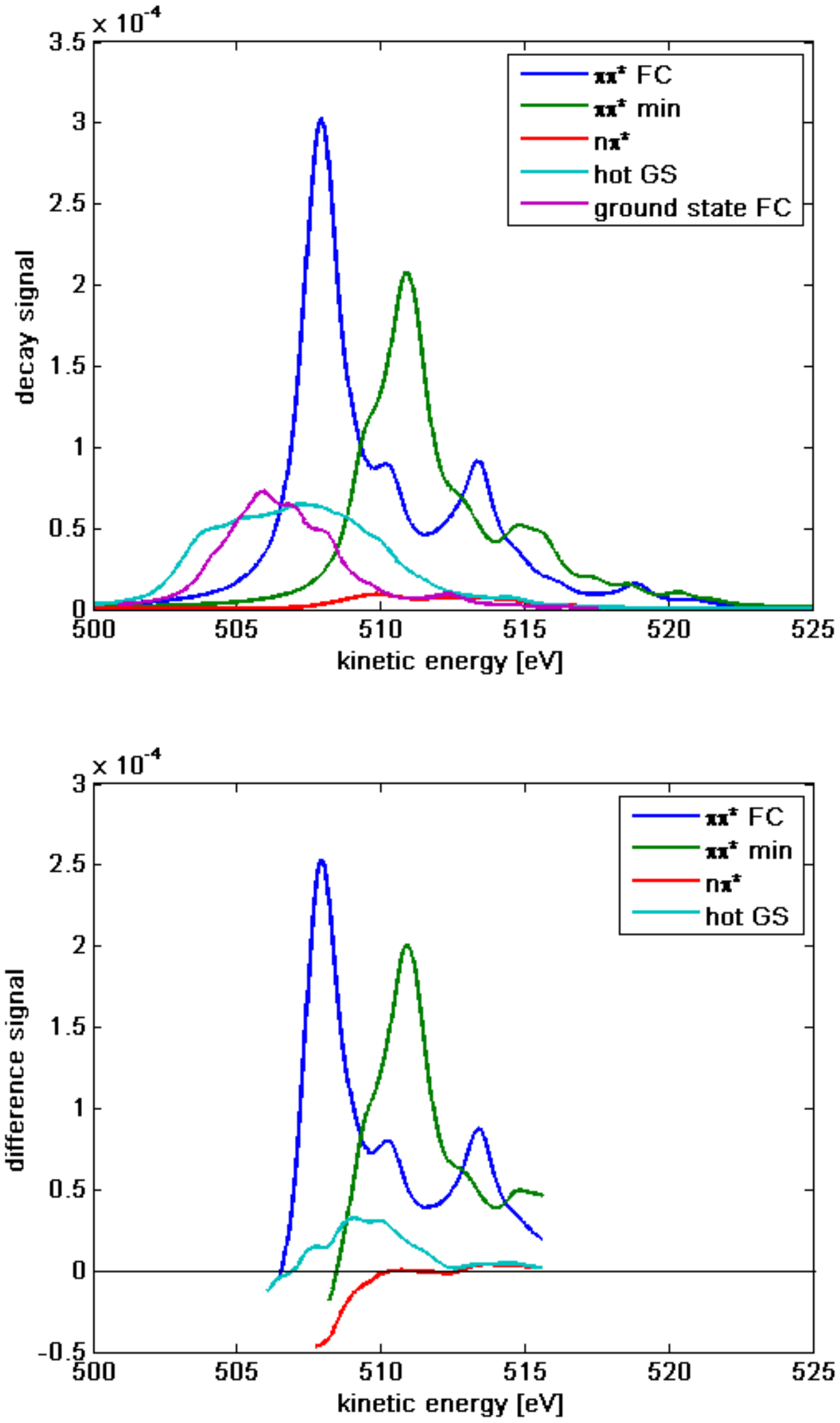

**Fig. 5:** a) Calculated Auger decay spectra for the GS, $\pi\pi^*$ and $n\pi^*$ at different internuclear geometries. The ground state Auger decay is calculated at the Franck-Condon geometry (purple) corresponding to the Auger decay of a non-UV excited molecule. The $\pi\pi^*$ Auger decay at the Frank-Condon geometry (green) reflects the wavepacket directly after and during the UV excitation. The $n\pi^*$ decay (red) is evaluated at an average of several vibrationally "hot" geometries, in a similar way to the hot ground state (light green).
b) Difference spectra between Auger decay of excited states and the ground state at Frank-Condon geometry. We observe mainly positive features in the selected spatial range (given by the calculation range for the state specific Auger rates). The only exception is the $n\pi^*$ state, which produces a negative signal below 510 eV kinetic energy. The simulation interval in the stick spectra (see Appendix 1) determines the kinetic energy range in the difference spectra.

Figure 5a shows the calculated Auger spectra for different photoexcited states and the ground state in its ground state minimum geometry. The calculations are only possible over a limited range of energies, as explained above. Generally, the UV excited states show far more Auger signal than the initial ground state at any given kinetic energy. The decay from strong Auger rates towards weak Auger rates with increasing kinetic energy, visible in the ground state spectra in Fig. 3, is shifted towards higher decay energies. This is due to the increased kinetic energy by relaxation from a core AND valence excited intermediate state. Furthermore the nuclear motion away from the Frank-Condon geometry leads to lower potential energies in the generally repulsive dicationic states and thus higher kinetic energies of the Auger electrons. As expected from the intuitive arguments used above, the $n\pi^*$ state forms an exception among the photoexcited states having a decay strength similar or even weaker than the initial ground state. The difference spectra in Fig. 5b reflect this behavior. The features in Fig. 5b show some resemblance with the experimental difference spectra in Fig. 4a. The $\pi\pi^*$ spectrum grows towards lower Auger energies, the fact that it is peaked at ~507 eV in perfect agreement with the experiment is not completely sound yet, since the energy range of our simulations is too limited. Further, currently unknown, simulated Auger lines towards the lower energy might redshift this peak. A similar statement is true for the $\pi\pi^*$ peak at the minimum geometry. The argument for a long lived hot ground state visible in our experimental data is strongly supported by the simulations, since the $n\pi^*$ would yield small or even negative difference signals in the selected energy range. The ratio of the simulated difference spectra of $\pi\pi^*$ Franck-Condon (blue in Fig.5b) and the hot ground state (light green in Fig.5b) is very similar to the observed ratio of 2:0.7 between the initial peak at 507 eV and the 20 ps later peak at ~511 eV in Fig. 4a. This confirms our interpretation in terms of initial $\pi\pi^*$ and late hot ground state population. *We thus confirm that we have directly measured the product state for nucleobase photoprotection for the first time in the isolated molecule. We estimate its filling time to be 300 fs, deduced from the* $\pi\pi^*$ *decay time and the final signal rise time.*

Our observations demonstrate the strengths of Auger spectroscopy. The core hole ionization at our photon energies is possible no matter in which electronic state the molecule resides. Ultrafast spectroscopy with ultraviolet pump and probe pulses generally loses the signal, since the ionization potential grows above the photon energy electronic relaxation occurs. We show that it is possible to observe the product of photoprotection; the hot ground state leaves its signature in our spectra. The element specificity of the core hole decay as well as the strong chemical sensitivity that is displayed in the Auger matrix element (Eq. 3 and 4) lead to new types of propensity rules. We showed that excited electronic states with reduced population in lone pair orbitals at the core hole location show a specifically weakened Auger decay, and thus a negative difference spectrum feature, at high kinetic Auger electron energies. This presents a real complementary picture compared to dipole selection rules used in fluorescence and absorption spectroscopy. The absence of an Auger decay from the $n\pi^*$ state allowed us to attribute the long lasting spectral features to the vibrationally hot electronic ground state. This clear interpretation would not have been so clear for spectrally congested cases where the npi* and electronic ground state contribute. We expect that the strongest $n\pi^*$ decay channels are shifted to lower kinetic energies out of the window between 500 and 520 eV (Appendix 1). The experimental spectra in Fig. 4 show a 500 fs rise between 485 and 495 eV. We cannot rule out, that this energy range is also congested with Auger transitions from the hot ground state, and therefore we currently abstain from further discussions of this channel.

We now compare the measured and interpreted time constants with the transients known from the gas phase thymine literature. We observe signal decays *and* delayed signal rise in our spectra. In contrast, the published isolated molecule spectra of all nucleobases contain only signal decays, which are fitted to several rise and decay times. This qualitative difference is remarkable and we attribute it to the fact that x-rays, in contrast to visible or UV probe pulses, can ionize the molecule independently of their nuclear geometry, thus we are able to also probe very low lying states. Furthermore, the Auger kinetic energy difference between the ground state and the pipi* state is large enough so that the filling of the vibrationally hot ground state becomes distinct from other channels visible. Very short decay constants on the order of 100 fs have been documented in the literature [Kang_2002,Ullrich_2004, Canuel_2005, Hudock_2007] and have been attributed as a loss of signal out of the Franck-Condon window in the $\pi\pi^*$ state. The 200 fs decay constant between 505 and 515 eV kinetic energy generally supports this interpretation and the difference in decay constant can simply be attributed to different probe window sizes in the nuclear coordinate space. The time constant of 500 fs attributed by us as the time to fill the ground state has been fitted before [Ullrich_2004] and interpreted as a relaxation into a minimum on the $\pi\pi^*$ state which then empties on a ~5 ps timescale. We observe a long lasting >20 ps signal, which rules out this interpretation for the transients described here, once more strengthening the interpretation of a vibrationally hot ground state.

## Conclusions

We have presented a new method called DUXAP for ultrafast spectroscopy of isolated molecules. The technique is particularly well suited to follow ultrafast electronic relaxation, since the high photon energy of the SXR probe pulse warrants for core ionization independent of the valence electron configuration. We have demonstrated a particular advantage of the DUXAP method, which lies in propensity rules allowing a different view on dynamics compared to dipole selection rules.

We have applied the method to the ultrafast electronic relaxation of photoexcited thymine, which is a topic of sufficient research and also sufficient controversy regarding the details of the electronic relaxation. Two relaxation paths have been proposed, one being a direct decay to the vibrationally hot electronic ground state, the other one being a relaxation into the $n\pi^*$ state. We observe the decay of transient population on the initial $\pi\pi^*$ with a time constant around 300 fs. The propensity for $n\pi^*$ observation with DUXAP is very low in the kinetic energy window of our study and thus we succeed to isolate the direct ground state decay mechanism for the first time in the ultrafast spectroscopy of isolated nucleobases. We estimate the hot ground state to be filled within 300 fs after the ultraviolet excitation.

We thank Wanli Wang and David Hardy for their help at Beamline 8 of the Advanced Light Source.

This work was supported by the AMOS program within the Chemical Sciences, Geosciences, and Biosciences Division of the Office of Basic Energy Sciences, Office of Science, U.S. Department of Energy. M. G. acknowledges funding via the Office of Science Early Career Research Program through the Office of Basic Energy Sciences, U.S. Department of Energy. Portions of this research were carried out at the Linac Coherent Light Source (LCLS) at the SLAC National Accelerator Laboratory. LCLS is an Office of Science User Facility operated for the U.S. Department of Energy Office of Science by Stanford University. Other portions of this research were carried out at the Advanced Light Source, which is supported by the Director, Office of Science, Office of Basic Energy Sciences, of the U.S. Department of Energy under Contract No. DE-AC02-05CH11231.

**APPENDIX 1 – Theory details**

We presented two methods for calculating the Auger decay. The ADC(2) method presents the most accurate standard, but it is computationally expensive and currently only applicable to the electronic ground state of the molecule. We have therefore implemented an alternative method based on the calculation of dicationic state energies and Auger matrix elements according to Eq. 3 and 4 above. We controlled the quality of the latter method by calculating the ground state Auger decay and comparing it to the ADC(2) method in the energy window presented in Fig. A1. The comparison shows reasonable agreement between the two methods, and we move on to simulate the Auger decay of the excited states. The decay rates (Γ) are scaled by a factor 0.1 compared to the main text, but otherwise completely identical to the decay strength presented in the main text.

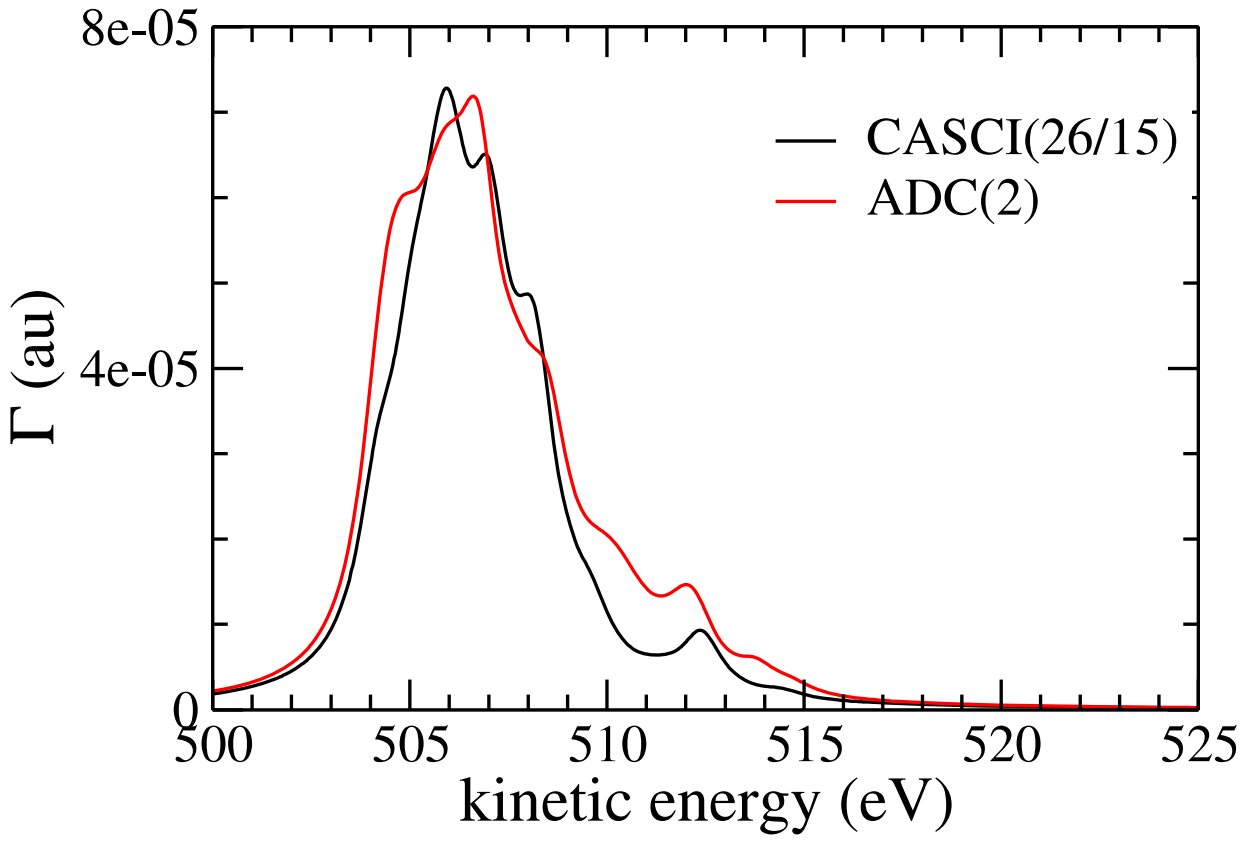

Fig. A1: Calculated Auger electron spectra of ground state thymine at Franck-Condon geometry. The spectra were computed using the CASCI(26/15)/6-311G* method and ADC(2)/cc-pVDZ method.

The Auger decay matrix elements of the electronic ground state in the Franck-Condon geometry are large when molecular orbitals 23a, 24a, 25a, 30a or 31a are involved. Here are the most dominant configurations that give rise to fast decay rates (note: MOs 20a~34a are shown):

| Config | Rate | 2 hole |
|---|---|---|
| 222222222202220 | 0.000024764331 | 30 |
| 222222222220220 | 0.000030634087 | 31 |
| 2222222222ab220 | 0.000053344170 | 30 31 |
| 222a222222b2220 | 0.000055103172 | 23 30 |
| 222a2222222b220 | 0.000068532818 | 23 31 |
| 2222a22222b2220 | 0.000010986320 | 24 30 |
| 2222a222222b220 | 0.000011921016 | 24 31 |
| 22222a2222b2220 | 0.000037827361 | 25 30 |
| 22222a22222b220 | 0.000046743610 | 25 31 |

In general, all these orbitals have very strong localization at the oxygen sites of thymine. MOs 24a, 25a, 30a and 31a are shown in the pipi* section. Here is 23a:

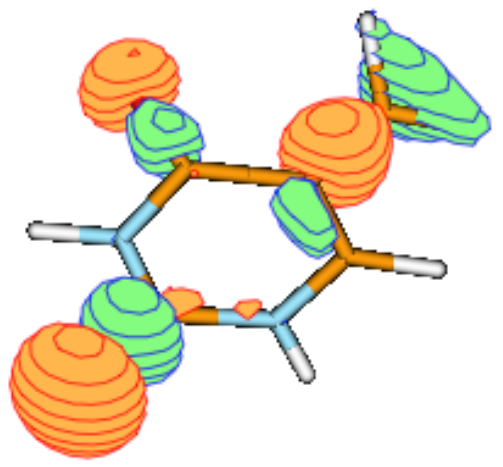

**Fig. A2:** MO 23a

**pipi* electronic state**

The Auger spectra of Thymine in its FC and $S_2$ minimum geometries are presented in the folowing. Note that we have labeled the most dominant peaks in the spectra.

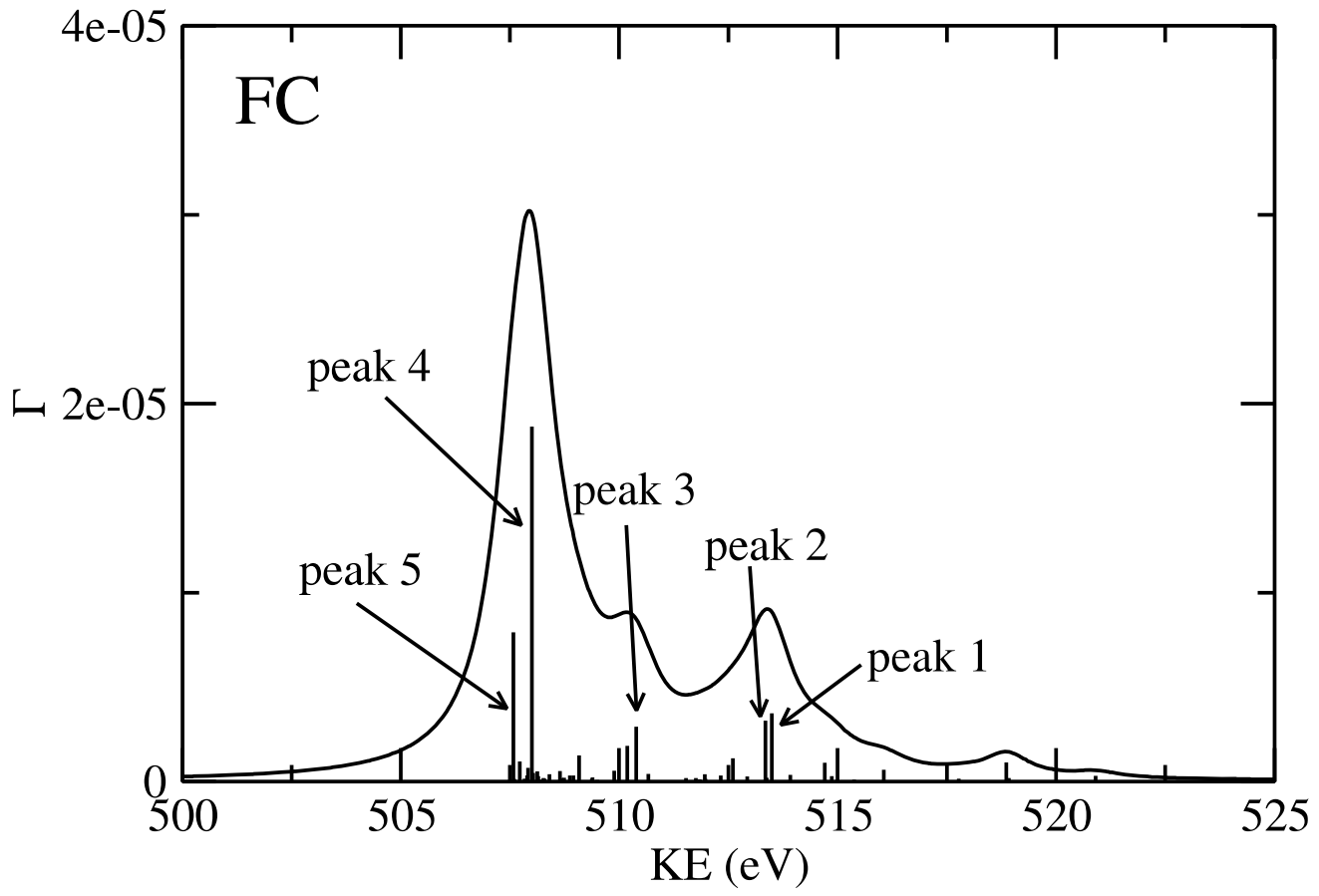

**Fig. A3:** Auger decay of the pipi* state for the Frank-Condon nuclear geometry.

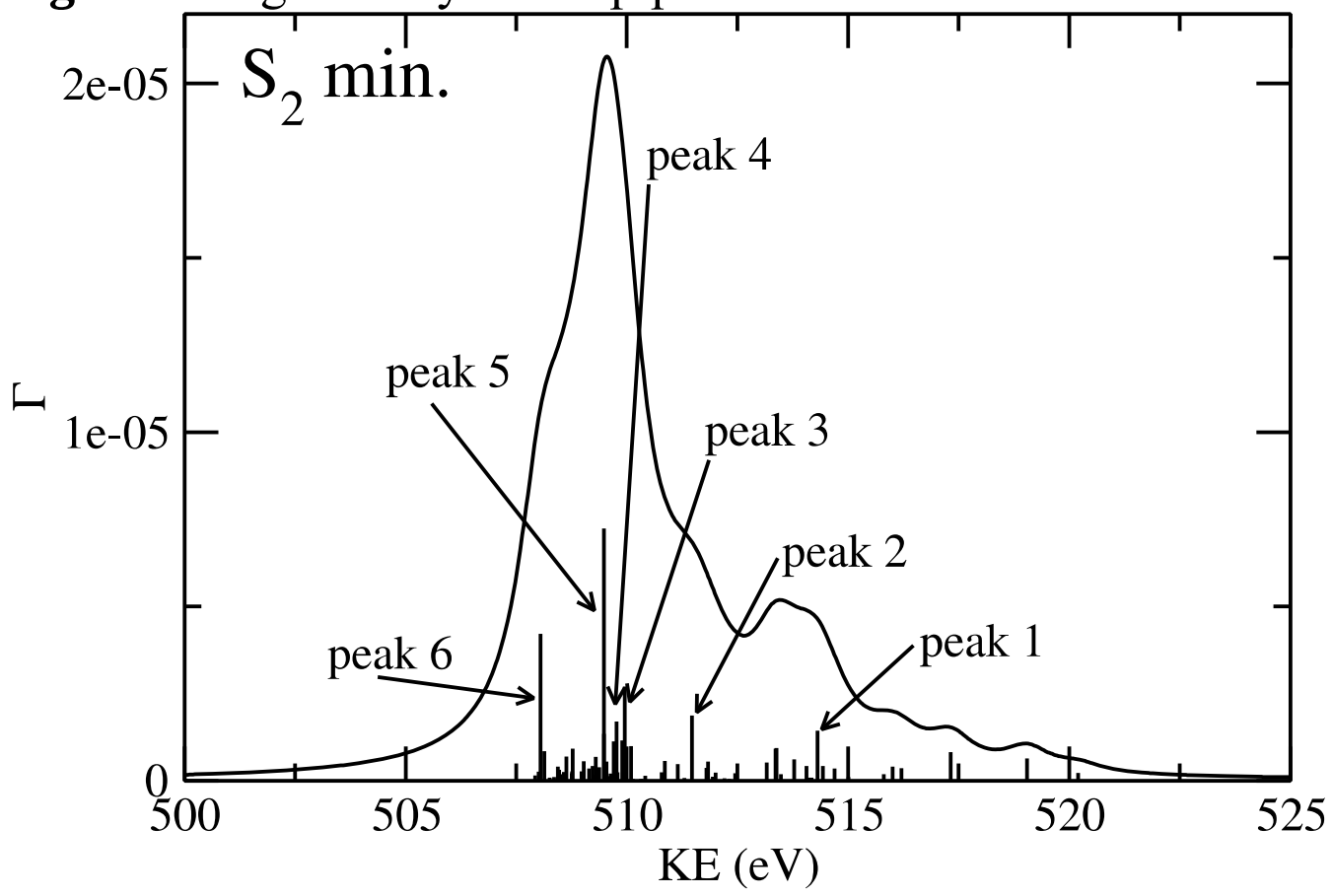

**Fig. A4:** Auger decay of the pipi* state for the minimum nuclear geometry.

The strongest pipi* state Auger decay involves molecular orbitals 24a, 25a, 30a or 31a.
Here are the dominant dication configurations that give rise to fast transition rates:

FC geometry:

| Config | Rate | 2 hole |
|---|---|---|
| 2222222222202ab | 0.0000303559020 | 31 |
| 22222a22222b2ab | 0.0000451447430 | 25 31 |
| 2222222222ab2ab | 0.0000530316690 | 30 31 |
| 22222a2222b22ab | 0.0000367440190 | 25 30 |
| 2222222222022ab | 0.0000246022640 | 30 |

The strongest Auger transitions in the pipi* minimum geometry [Hudock_2007] are given below:

| Config | Rate | 2 hole |
|---|---|---|
| --------- | ------- | ------ |
| 2222222222a2bab | 0.0000138478440 | 30 32 |
| 222222222a2b2ab | 0.0000234705860 | 29 31 |
| 2222222222ab2ab | 0.0000118499530 | 30 31 |
| 2222222222202ab | 0.0000628062810 | 31 |
| 2222a222222b2ab | 0.0000747272340 | 24 31 |
| 22222a22222b2ab | 0.0000380777580 | 25 31 |

The MOs 31 and 30 are oxygen lone pair orbitals, the lower MOs 24 and 25 show again strong oxygen site localization.

24a 25a 30a 31a

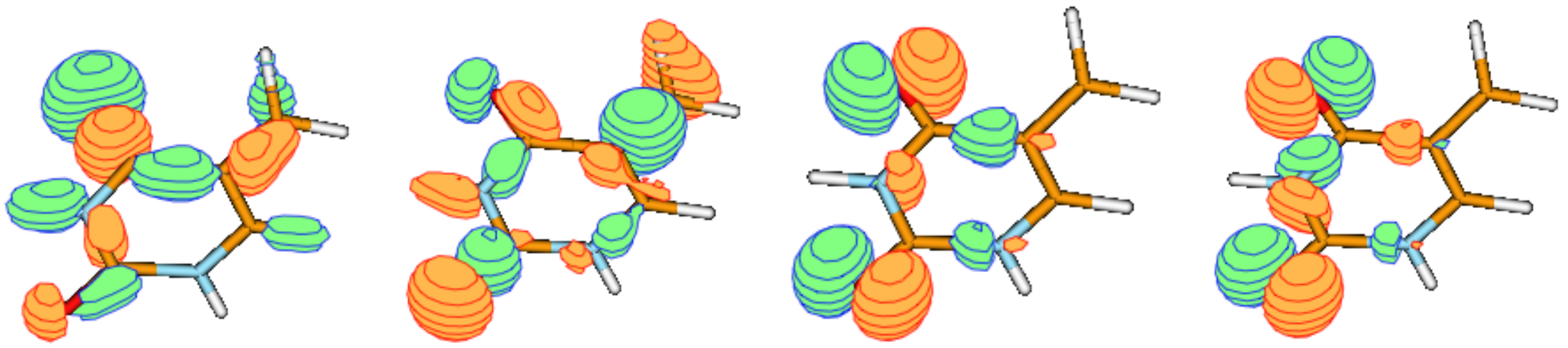

**Fig. A5:** MOs participating strongly in ground state and pipi* state Auger decay.

**npi* electronic state**

The npi* spectrum in the main text (see Fig. 5) has been averaged over several nuclear geometries., Here we present only one spectrum at the conical intersection with the pipi* state geometry.

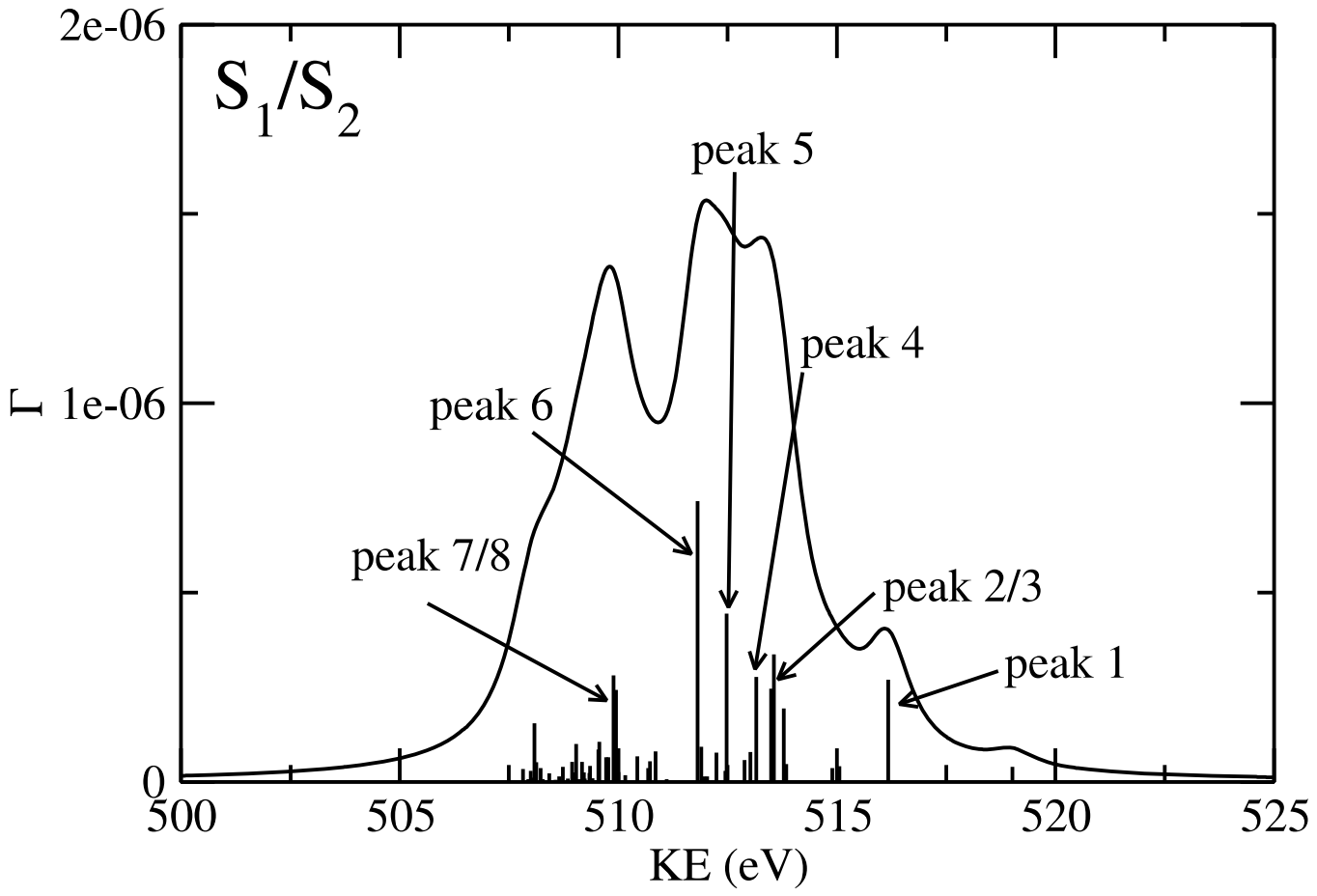

**Fig. A6:** Auger decay of the npi* state for the conical intersection npi*/pipi* nuclear geometry.

In this case, the Auger rates are more than one order of magnitude reduced compared to the obove mentioned from the ground state and pipi*.  The strongest Auger decay now involves orbitals 29a, 30a 32a or 33a:

29a 30a 32a 33a

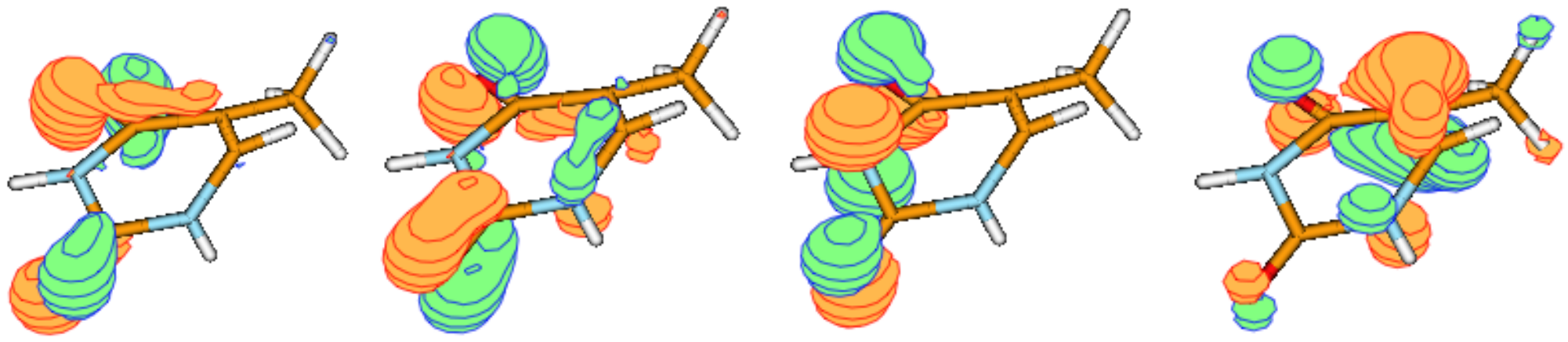

**Fig. A7:** MOs participating in the npi* Auger decay in the energy window shown in the previous figure.

-The dominant dication configurations that give rise to fast transition rates are

| Config | Rate | 2 hole |
|---|---|---|
| 22222222222ab20 | 0.0000015106220 | 32 34 |
| 2222222222ab220 | 0.0000017595830 | 30 34 |
| 22222222222aabb | 0.0000022170450 | 32 33 |
| 2222222222aa2bb | 0.0000026206070 | 30 33 |
| 222222222aba22b | 0.0000051412250 | 29 30 |
| 222222222a2ab2b | 0.0000037753620 | 29 32 |
| 2222222222202ab | 0.0000010763210 | 31 33 |

We observe that, in contrast to the strong ground state and pipi* decays, the pi* (34) and  pi (33) orbitals are involved. Thus it is clear that these Auger transitions must be weak since those orbitals are highly delocalized over the entire molecule.

We note, however, that there are other dication configurations that can give rise to even larger rates, which should appear at lower KE. We can calculate the matrix elements involving MOs 23a, 24a, 25a or 31a as given below, but have not yet succeeded in calculating the energies of the respective dicationic states.

| Config | Rate | 2 hole |
|---|---|---|
| 222a2222222022b | 0.000026476304 | 23 31 |
| 2222a222222022b | 0.000051828349 | 24 31 |
| 22222a22222022b | 0.000036206987 | 25 31 |
| 222222222a2022b | 0.000025668279 | 29 31 |
| 22220222222a22b | 0.000016985508 | 24 |

-Here are the MOs

23a 24a 25a 31a

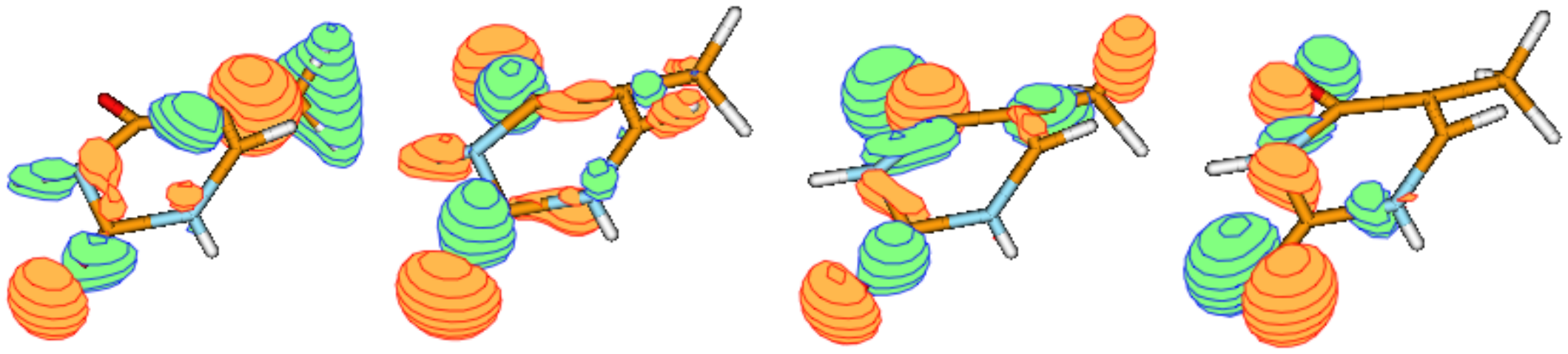

**Fig. A8:** MOs contributing to strong Auger features at energies, not yet determined.

We can however intuitively understand why these dicationic states must ly much higher in energy, resulting in lower kinetic energy Auger decay features. The dicationic states are characterized by three holes in the vicinity of the oxygen core hole: two holes in the shallow bound n orbital, and one in a deeper bound valence orbital strongly localized at oxygen. First of all, the missing screening resulting from the deeper valence orbital will shift the higher n-orbitals to lower binding energies. This results in a lowering of the Auger decay energy.

A further second argument results from the inspection of the ADC(2) calculation details.

We looked at simpler 2h configurations computed with the ground state orbitals. What we want to compare, therefore, are the energies of (n,pi) holes with the corresponding (n,n) holes. There are two principal n orbitals that give the largest contributions to the oxygen Auger spectrum and they are orbitals 30 and 31. For orbital 30 the (n,n) configuration lies about 9 eV above the (n,pi) one. About half of this gap is due to orbital energy difference and the other half is due to the larger repulsion in the (n,n) state (12 eV vs 6 eV). For orbital 31, (n,n) is 8 eV above (n,pi) of which again 6 eV are due to larger repulsion. For comparison, the repulsion in the more delocalized (pi,pi) state is about 8 eV.

These two considerations support the surmise that the npi* decay involving THREE orbitals strongly localized at oxygen may be at over 15 eV lower kinetic energy than the pipi* decay. This would indeed result in decay patterns of the npi* state below 500 eV kinetic energy.